\documentclass[prb,twocolumn]{revtex4}
\usepackage[dvips]{graphicx}
\usepackage{latexsym,amsmath,amssymb,bm,euscript}

\def\etal{~\textit{et~al.}} 
\def\ra{\rangle} 
\def\la{\langle} 
\def\Hc{{\rm h.c.}}
\def\ET{$\kappa$-(ET)$_2$Cu$_2$(CN)$_3$~}

\begin{document}

\title{Variational study of triangular lattice spin-1/2
model with ring exchanges and \\
spin liquid state in $\kappa$-(ET)$_2$Cu$_2$(CN)$_3$}
\author{Olexei I. Motrunich}
\affiliation{Kavli Institute for Theoretical Physics, 
University of California, Santa Barbara, CA 93106-4030}

\date{December 20, 2004}

\begin{abstract}
We study triangular lattice spin-1/2 system with antiferromagnetic
Heisenberg and ring exchanges using variational approach 
focusing on possible realization of spin liquid states.
Trial spin liquid wave functions are obtained by Gutzwiller projection 
of fermionic mean field states and their energetics is compared against 
magnetically ordered trial states.
We find that in a range of the ring exchange coupling upon
destroying the antiferromagnetic order, the best such spin
liquid state is essentially a Gutzwiller-projected Fermi sea
state.  We propose this spin liquid with spinon Fermi surface as a 
candidate for the nonmagnetic insulating phase observed in the 
organic compound $\kappa$-(ET)$_2$Cu$_2$(CN)$_3$, 
and describe some experimental consequences of this proposal.
\end{abstract}

\maketitle

\section{Introduction}
This paper reports a variational study of spin-1/2 Heisenberg 
antiferromagnet with ring exchanges on a triangular lattice.
One motivation for this study is the exact diagonalization work of 
LiMing\etal\cite{LiMing} and Misguich\etal\cite{Misguich} on this 
system proposing that it realizes spin liquid states.  
We are particularly interested in spin liquids that may occur near the 
Heisenberg antiferromagnetic state.
Multiple-electron exchanges are believed to be important near 
quantum melting and metal-insulator transitions.
The specific model considered here may also be relevant for the 
description of a tentative spin liquid state observed in the 
quasi-two-dimensional organic compound
$\kappa$-(ET)$_2$Cu$_2$(CN)$_3$,\cite{Kanoda} which is close to
metal-insulator transition.  
Imada\etal\cite{Imada} studied appropriate Hubbard model on the 
triangular lattice and found an insulating regime with no spin order.
The ring exchange spin model can be viewed as derived from the Hubbard
model by a projective transformation, which is appropriate in the 
presence of the charge gap.

The present work attempts to understand possible spin liquid states
in the ring exchange model by examining candidate ground state 
wave functions.  This is complementary to the exact diagonalization 
studies, since knowing the character of a candidate wave function
can give significant intuition.

The model Hamiltonian on the triangular lattice is,
in the notation borrowed from Ref.~\onlinecite{Misguich},
\begin{equation}
\label{Hring}
{\hat H}_{\rm ring} = 
J_2 \sum_{
\begin{picture}(17,10)(-2,-2)
	\put (0,0) {\line (1,0) {12}}
	\put (0,0) {\circle*{5}}
	\put (12,0) {\circle*{5}}
\end{picture}
} P_{12}
+ J_4 \sum_{
\begin{picture}(26,15)(-2,-2)
        \put (0,0) {\line (1,0) {12}}
        \put (6,10) {\line (1,0) {12}}
        \put (0,0) {\line (3,5) {6}}
        \put (12,0) {\line (3,5) {6}}
        \put (6,10) {\circle*{5}}
        \put (18,10) {\circle*{5}}
        \put (0,0) {\circle*{5}}
        \put (12,0) {\circle*{5}}
\end{picture}
} \left( P_{1234}+P_{1234}^\dagger \right) ~.
\end{equation}
The two-spin exchanges are between all nearest neighbors
and reduce simply to Heisenberg interactions,
$P_{12} = P_{12}^\dagger = 2 {\bm S}_1 \cdot {\bm S}_2 + \frac{1}{2}$.
The four-spin ``ring exchanges'' are around all rhombi
of the triangular lattice.

\begin{figure}
\centerline{\includegraphics[width=\columnwidth]{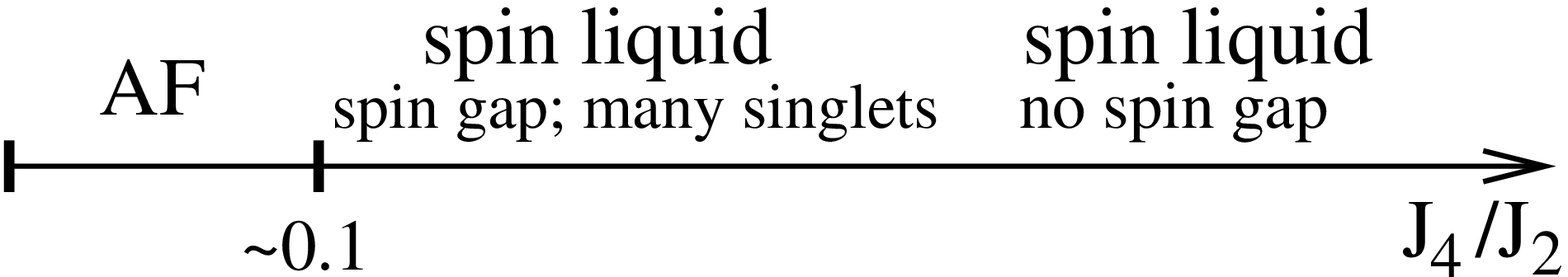}}
\vskip -2mm
\caption{Phase diagram of the model Eq.~(\ref{Hring}) from exact
diagonalization study of Ref.~\onlinecite{LiMing, Misguich}.
The magnetic order is destroyed for $J_4 \gtrsim 0.07-0.1$; 
a spin gap is observed in the regime 
$0.1 \lesssim J_4 \lesssim 0.25$, but also many singlets below the 
spin gap.  The spin gap is decreasing for $J_4 \gtrsim 0.175$.}
\label{ediag_phased}
\vskip 2mm

\centerline{\includegraphics[width=\columnwidth]{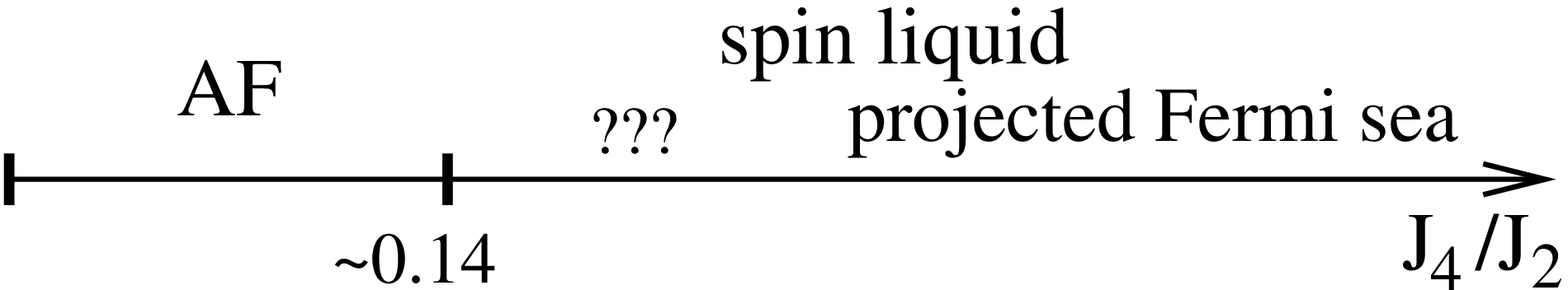}}
\vskip -2mm
\caption{Variational phase diagram for the Hamiltonian Eq.~(\ref{Hring}).
The AF ordered variational state has the lowest energy for small
$J_4$, but becomes unstable for $J_4 \gtrsim 0.14$ compared with 
the fermionic spin liquid states.  One example of such spin liquid is
the projected $d_{x^2-y^2} + i d_{xy}$ superconductor ansatz, 
with the optimal variational parameter 
$(\Delta/t)_{\rm var} = 0.22, 0.13, 0.05, 0.02$ for
$J_4 = 0.15, 0.20, 0.25, 0.30$ respectively.
Some other ansatze give very close optimal energy, and the situation is 
particularly not clear near the AF state.  But for $J_4 \gtrsim 0.3$, 
our best ansatze become essentially the projected Fermi sea state.
We caution that for significantly larger $J_4$ states with more 
complicated magnetic orders -- e.g. with four-sublattice order -- may 
enter the energetics competition,\cite{four_sublat} which is not 
considered here.
}
\label{var_phased}
\end{figure}

In the following, we consider only antiferromagnetic coupling $J_2>0$ 
and positive $J_4 \geq 0$; for brevity, we set $J_2=1$.  
When $J_4=0$, the system is the familiar Heisenberg antiferromagnet 
on the triangular lattice and has a three-sublattice antiferromagnetic 
(AF) order.
Exact diagonalization study of Ref.~\onlinecite{LiMing} proposes
the phase diagram summarized in Fig.~\ref{ediag_phased}.
The AF order is preserved for small $J_4 \lesssim 0.07-0.1$,
but is destroyed for larger $J_4$ and a spin gap opens up.
However, in the regime $0.1 \lesssim J_4 \lesssim 0.25$ reported in 
Ref.~\onlinecite{LiMing}, there are apparently many singlet excitations 
below the spin gap.  Also, the spin gap starts to decrease for
$J_4 \gtrsim 0.175$.

In the exact diagonalization studies, it is hard to say which physical 
state is realized in the absence of clear signatures of some particular 
phase.  The question of possible spin liquid states is taken up here 
by considering variational spin liquid wave functions on the triangular 
lattice.  Specifically, we consider one family of such states obtained by
Gutzwiller-projecting singlet fermionic mean field states.
\cite{WenPSG, ZhouWen}
We determine the result of the competition with the AF ordered state 
by comparing against variational wave functions with long range 
magnetic order.\cite{HuseElser}

The result of the variational study is summarized in 
Fig.~\ref{var_phased}.
For small $J_4 \lesssim 0.14$, the AF state is stable compared with 
the tried spin liquid states.  For larger $J_4$, we find spin liquid
states that have lower variational energy than the magnetically ordered
state.  
For example, we find that projected superconductor ansatze perform
well in the regime $0.14 \lesssim J_4 \lesssim 0.3$.  
More specifically, ansatze with anisotropic extended $s$-wave, 
$d_{x^2-y^2}$, and $d_{x^2-y^2} + i d_{xy}$ pairing patterns 
have very close optimal energies and much lower than the energy of the 
trial AF state.  Unfortunately, we conclude that the present study is 
not sufficient to address the nature of the spin liquid in this regime,
which we indicate with question marks in the figure.
Our observation that the improvement in the trial energy is little 
sensitive to the specific pairing pattern may be an indication 
that the present restricted study cannot access the correct ground state 
in this regime.

A more robust conclusion from our study of such spin liquids
is that the best ansatze are close to the projected Fermi sea state 
and become more so for increasing $J_4$.
Thus, for $J_4 \gtrsim 0.3-0.35$ the variational $\Delta$ in our ansatze
reduces to essentially zero (below few percent of the hopping amplitude),
and the ground state is essentially the projected Fermi sea.

The aptitude of the projected Fermi sea state can be intuitively 
understood as follows.  We can view the ring exchange term with 
positive $J_4$ as arising from the electron hopping in the underlying 
Hubbard model (which we assume is in the insulating phase).
Therefore, such ring exchange $J_4>0$ wants the fermions to be as 
``delocalized'' as possible, and this ``kinetic energy'' is best 
satisfied in the simple hopping ansatz.  A more formal mean field
argument is given in Sec.~\ref{sec:fermSUN}

\vskip 2mm
We now discuss the indications of this study for the possible
spin liquid state in $\kappa$-(ET)$_2$Cu$_2$(CN)$_3$.
This material is close to the metal-insulator transition,
so the role of the electron kinetic energy is clearly important.  
Based on the experience with the ring exchange model, we therefore 
propose that the projected Fermi sea state is a good candidate 
ground state close to the metallic phase.
We verify this more explicitly by considering a model with
ring exchanges obtained by a projective transformation of the 
triangular lattice Hubbard model at order $t^4/U^3$.
For the \ET compound we estimate $J_4/J_2 \simeq 0.3$.
The results are summarized in Fig.~\ref{material_phased}.
The work of Ref.~\onlinecite{Imada} on the triangular lattice
Hubbard model can be interpreted as an elaborate numerical study 
building up on free-fermion states, and hints some support to 
the proposed projected Fermi sea phase.

The proposed picture has many physical consequences.  
We have a Fermi surface of spinons, and therefore expect no spin gap and 
finite spin susceptibility down to zero temperature consistent with the 
experimental observations.
An accurate treatment of the no-double-occupancy constraint and 
fluctuations requires that the spinons are coupled to a fluctuating 
$U(1)$ gauge field.  Such spinon-gauge field system has been studied 
extensively and is expected to exhibit some unusual behavior.
\cite{Reizer, PALee, Polchinsky, Altshuler, Nayak, YBKim, Senthil}
For example, one expects a singular contribution to the specific heat 
$C_{\rm sing} \sim T^{2/3}$ at low temperatures in two dimensions;
the corresponding enhancement in ``spin entropy'' has concrete
consequences for the phase boundaries.

\vskip 2mm
The rest of the paper is organized as follows.
In Sec.~\ref{sec:states} we specify the variational states considered
in this work.  In Sec.~\ref{sec:fermSUN} we seek qualitative 
understanding of the ring exchange energetics by considering
a fermionic large-$N$ treatment of the ring exchange Hamiltonian.
In Sec.~\ref{sec:exp} we consider the connection with the 
triangular lattice Hubbard model and possible application to 
$\kappa$-(ET)$_2$Cu$_2$(CN)$_3$.  In particular, we describe 
experimental signatures of the proposed spinon Fermi surface - gauge 
system.

\section{Variational states and energetics}
\label{sec:states}
In this Section, we describe variational states used in the 
present work.
Trial spin liquid states are constructed by Gutzwiller projection of 
fermionic mean field states.\cite{WenPSG}
We compare their energetics against AF ordered trial wave functions 
constructed using the approach of Huse and Elser.\cite{HuseElser}

\vskip 2mm
{\bf Spin liquid trial states:}
The starting point here is the fermionic mean field treatment of the 
Heisenberg model.  A recent and very detailed description can be found in
Ref.~\onlinecite{WenPSG}.
The setup for constructing trial wave functions is as follows.
Each spin operator is written in terms of two fermions
$c_{r\uparrow}$ and $c_{r\downarrow}$,
${\bm S}_r = c_r^\dagger \frac{\bm \sigma}{2} c_r$, 
with precisely one fermion per site.
Heisenberg exchange interaction is written as a four-fermion 
interaction, which is then decoupled in the singlet channel.  
A convenient formulation of the mean field is to consider 
general spin rotation invariant trial Hamiltonian
\begin{equation}
H_{\text{trial}} = -\sum_{rr'} 
\left[
t_{rr'} c_{r\sigma}^\dagger c_{r'\sigma}
+ (\Delta_{rr'} c_{r\uparrow}^\dagger c_{r'\downarrow}^\dagger + \Hc) 
\right] ~, 
\end{equation}
with $t_{r'r}=t_{rr'}^*$, $\Delta_{r'r}=\Delta_{rr'}$.
For each such trial Hamiltonian we obtain the corresponding
ground state.  An $SU(2)$ invariant formulation of the
single occupancy constraint is that the isospin operator
${\bm T}_r \equiv \psi_r^\dagger \frac{\bm \tau}{2} \psi_r$
is zero on each site; here $\psi_{r\uparrow} = c_{r\uparrow}$,
$\psi_{r\downarrow} = c_{r\downarrow}^\dagger$, 
and $\tau^{1,2,3}$ are Pauli matrices.  In the mean field, 
we require that this constraint is satisfied on average,
which is achieved by tuning appropriate on-site terms. 
Going beyond the mean field, the physical spin wave function is 
obtained by projecting out double occupation of sites.

Many such trial states can be constructed, but there is also
a gauge redundancy in this construction.
Here, one is helped considerably by the recently available 
classification scheme of X.-G.~Wen\cite{WenPSG, ZhouWen} that allows 
one to construct all possible such fermionic mean field states that 
lead to physically distinct spin liquids with specified lattice 
symmetries.

We numerically evaluate the expectation values of the two-spin and 
four-spin exchanges in such states using standard determinantal
wave function techniques (so-called variational Monte Carlo).\cite{vmc}
We consider ansatze with different sets of lattice symmetries,
with and without time reversal, but primarily we focus on the 
nearest-neighbor ansatze that respect upon projection the lattice 
translation symmetry.
We then vary the parameters to optimize the trial energy.

\vskip 2mm
{\bf AF ordered trial states:}
We want to compare the ring exchange energetics of the spin liquid 
states with the energetics of the antiferromagnetically ordered states.  
For this purpose, we use the family of variational states considered by 
Huse and Elser,\cite{HuseElser} which capture well the Heisenberg
model energetics.
Our primary goal here is to see how the AF state is disfavored by 
the ring exchanges.  Since we are comparing with rather different
states and are looking for the energy level crossing, we do not need to 
know the ground state energy very accurately, and the wave functions of 
Ref.~\onlinecite{HuseElser} should be sufficient to get rough idea of the
ring exchange energetics in the AF state.
For details on these wave functions and numerical evaluations, 
the reader is referred to the original paper.

\vskip 2mm
{\bf Variational results:}
We compared the trial energies of the AF ordered states and the 
fermionic spin liquid states, and the result is summarized in 
Fig.~\ref{var_phased}.
For small $J_4$, the ordered states have lower energy, but for
$J_4 \gtrsim 0.14$ the spin liquid states win.
The optimal spin liquid ansatze have the following structure.  
The dominant part is the uniform triangular lattice hopping $t_{rr'}$, 
and for $J_4 \gtrsim 0.3-0.35$ we essentially find the projected 
Fermi sea state.  In the intermediate regime 
$0.14 \lesssim J_4 \lesssim 0.3$, we find that the trial energy is 
improved upon adding $\Delta_{rr'}$ correlations into the mean field 
wave function.  Somewhat perplexingly, we find that the result is 
not very sensitive to the specific ``pairing'' pattern.  
Thus, optimized wave functions with extended anisotropic $s$-wave, 
$d_{x^2-y^2}$, and $d_{x^2-y^2} + i d_{xy}$ pairing patterns have 
close energies.
This may be an indication of an instability towards a state
that cannot be captured in the context of the trial fermionic
states.  The situation is particularly inconclusive close to the
AF phase, where several other trial states have competitive energies.

In summary, we find that ring exchanges disfavor the AF ordered state 
compared with the fermionic spin liquid states, but our study is 
not conclusive as to which spin liquid state is realized when the 
transition happens.
Away from the transition, we suggest that the optimal spin liquid
state is the projected Fermi sea state.

\section{Fermionic large $N$ study of the ring exchange energetics}
\label{sec:fermSUN}
In this Section, we present a fermionic large $N$ study of the ring 
exchange Hamiltonian.  Here, natural ``trial'' states are pure 
hopping states, and this approach gives us some insight 
into their energetics.  In particular, it shows how the ring exchanges 
favor the uniform hopping state, i.e. the projected Fermi sea state.  
The treatment below was suggested to the present author by T.~Senthil.

Consider the following generalization of the ring exchange
Hamiltonian Eq.~(\ref{Hring}) to an $SU(N)$ spin model
\begin{eqnarray*}
& \hat H_{SU(N)} & = \frac{J}{N} \sum_{\la 12 \ra}
(c_{1\alpha}^\dagger c_{1\beta}) (c_{2\beta}^\dagger c_{2\alpha}) \\
&+& \frac{K}{N^3} \sum_{P}
\left[
(c_{1\alpha}^\dagger c_{1\beta}) (c_{2\beta}^\dagger c_{2\gamma})
(c_{3\gamma}^\dagger c_{3\delta}) (c_{4\delta}^\dagger c_{4\alpha})
+ \Hc \right]
\end{eqnarray*}
We use conventional fermionic representation with $N$ fermion flavors;
spin states on each site are viewed as states of $N/2$ fermions,
i.e. we have occupancy constraint
\begin{equation}
c_{r \alpha}^\dagger c_{r \alpha} = N/2
\end{equation}
for each site $r$.  In the above, summation over repeated flavor 
indices is implied.  
Our generalization of the exchange operators preserves the character of 
moving spins around a ring.  For $N=2$, this Hamiltonian 
reduces precisely to the spin-1/2 Hamiltonian Eq.~(\ref{Hring}) with
\begin{equation}
J=2 J_2, \quad K=8 J_4.
\end{equation}
A similar large $N$ formulation was considered in a different context in 
Ref.~\onlinecite{Wen:origin}.
We also remark here that the general $N$ formulation allows nontrivial 
exchanges involving three spins.  This is unlike the $N=2$ case where 
such three-spin exchange reduces to a combination of two-spin 
exchanges.  The three-spin exchanges can be easily included in the 
following analysis; to stay in line with the rest of the paper, 
we only consider the two-spin and four-spin exchanges.

We formulate the large $N$ procedure in the spirit of the variational 
approach.  Consider a single-particle `trial' Hamiltonian
\begin{eqnarray*}
\hat H_{\rm trial} &=&  
-\sum_{\la rr' \ra} \left( t_{rr'} c_{r\alpha}^\dagger c_{r'\alpha}
                          + \Hc \right)
-\sum_r \mu_r c_{r\alpha}^\dagger c_{r\alpha} .
\end{eqnarray*}
We find the ground state and use it as a trial wave function for
the Hamiltonian $\hat H_{SU(N)}$.  In the mean field, the occupancy 
constraints are implemented on average by tuning the chemical 
potentials $\mu_r$.  The trial energy to leading order in
$1/N$ is given by
\begin{eqnarray*}
\frac{E_{\rm mf}}{N} &=& -J \sum_{\la 12 \ra} |\chi_{12}|^2 
- K \sum_{P}
\left( \chi_{12} \chi_{23} \chi_{34} \chi_{41} + c.c. \right),
\end{eqnarray*}
where $\chi_{rr'}^{*} \equiv \la c_r^\dagger c_{r'} \ra$ is the
single-species expectation value.

We now have to minimize $E_{\rm mf}$ over the possible $t_{rr'}$ in the 
trial Hamiltonian.  This leads to the following self-consistency
conditions
\begin{equation}
\Lambda^{-1} t_{rr'} = J\chi_{rr'} 
+ \sum_{P=[1234]=[rr'34]} K \chi_{23}^* \chi_{34}^* \chi_{41}^* ~, 
\end{equation}
where the last sum is over all ring exchange plackets that 
contain the bond $\la rr' \ra$ as one of the consecutive bonds.
Also, we have explicitly indicated the fact that the trial 
energy does not depend on the absolute scale in the trial Hamiltonian
but only on the relative pattern of $t_{rr'}$.

We first make some general observations about this procedure.  
First of all, note that the self-consistency conditions
imply that the optimal state can have non-zero $t_{rr'}$
only on the bonds that have non-zero $J_{rr'}$ or that appear in some 
ring exchange placket. 
For the triangular lattice model studied here, we then have to consider 
only nearest-neighbor $t_{rr'}$.  
Second, we see quite generally that the ring exchange contribution
for a given placket has the form $-K|\chi|^4 \cos(\Phi_P)$,
where $|\chi|$ is the geometric mean of the absolute values
of $\chi_{rr'}$ around the placket, while $\Phi_P$ is the ``flux''
of the corresponding phase factors.  Thus, the positive ring exchange
wants to smear the fermions over the lattice with no fluxes.

To be more precise, let us consider several simple trial states.
\underline{Uniform flux} state has flux $\phi$ through each 
triangle.  The expectation values 
$\chi_{rr'} = \la c_{r'}^\dagger c_r \ra $ have the same pattern of 
fluxes as the input $t_{rr'}$, and the trial energy per site is
\begin{equation}
E_{\phi} = -3 J |\chi_\phi|^2 - 6 K |\chi_\phi|^4 \cos(2\phi) ~,
\end{equation}
since the flux through each rhombus is $2\phi$.
Among such flux states, we find that for $K \lesssim 2.76 J$ the best 
state has \underline{$\pi/2$ flux} through each placket 
(this state has the largest $|\chi_\phi|$), while for 
$K \gtrsim 2.76 J$ the best state has \underline{zero flux}  
The numerical values of the energy per site in the two states
can be obtained from
\begin{eqnarray}
E_{\phi=\pi/2} &=& -0.120 J + 0.0096 K, \\
E_{\phi=0} &=& -0.081 J - 0.0044 K ~.
\end{eqnarray}
For large enough $K/J$ the zero-flux state is stable against 
adding small flux $\phi$ because 
$|\chi_\phi|^2 / |\chi_0|^2 \lessapprox 1 + 0.2 \phi^2$.

We also considered so called \underline{dimer} states such that 
non-zero $t_{rr'}$ form nonoverlapping dimer covering of the lattice.  
These states break translational invariance, and any dimer covering
produces such a state.  It is well known that these states can have 
lower Heisenberg exchange energy in the large $N$ limit.  
This is because the occupied bonds attain the maximal expectation 
value $|\chi_{rr'}|_{\rm max}$ and their contribution can be 
sufficient to produce the lowest total energy.  
The energy per site in any dimer state is
\begin{equation}
E_{\rm dimer} = -0.125 J ~,
\end{equation}
and is indeed the lowest energy for $K=0$.  However, the dimer
states gain no ring exchange energy, and for $K \gtrsim 9.9 J$ the 
zero flux state becomes the lowest energy state.
Finally, the so called \underline{box} states have identical two-spin 
exchange energy with the dimer states but also nontrivial
fluxes and therefore do not enter the competition for the 
ground state for $K>0$.

\begin{figure}
\centerline{\includegraphics[width=\columnwidth]{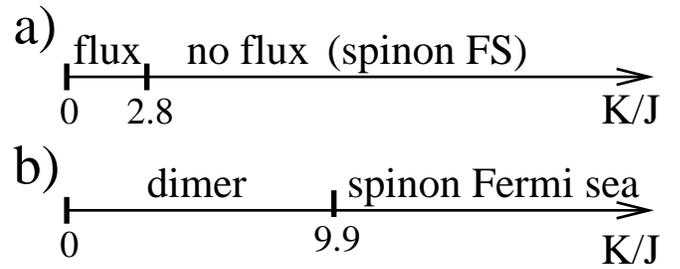}}
\vskip -2mm
\caption{Summary of the large $N$ study of the Hamiltonian
$\hat H_{SU(N)}$.  
a) Phase diagram from the mean field energy optimization over 
translationally invariant states. 
b) Full optimization. }
\label{largeN_phased}
\end{figure}

We performed full optimization over $t_{rr'}$ of the mean
field energy considering possible unit cells with up to four sites, 
and found that the above simple states are indeed sufficient to 
describe the ground state in the large $N$ limit:  
The optimal state is one of the dimer states for $K \lesssim 9.9 J$ 
and becomes the zero flux state for larger $K$.
The complete study is summarized in Fig.~\ref{largeN_phased}.

To make better connection with the spin-1/2 system, we remark that 
we expect the dimer states to be energetically disfavored even for 
small $K$ in the spin-1/2 case, e.g. compared with the flux states 
discussed above.  This is because the Gutzwiller projection enhances 
local antiferromagnetic correlations more strongly in the 
translationally invariant mean field states than in the 
dimerized states.\cite{Zhang}
More quantitatively, the enhancement factor for the Heisenberg energy is 
roughly $g_J^{\rm transl.inv.}=4$ for the translationally invariant 
states, while it is only $g_J^{\rm dimer}=2$ for the dimer states.  
Furthermore, we expect even stronger enhancements in the ring exchange 
energy upon the projection.  Taking all this into account, we expect the 
Fermi sea state to be favored for rather moderate $J_4/J_2$ in the 
spin-1/2 system.

To conclude the mean field discussion, our main message is that the 
positive ring exchange dislikes the fluxes and wants to make the system 
as uniform as possible.  This is best realized in the projected 
Fermi sea state.

\vskip 2mm
Going beyond the mean field, we obtain a theory of fermions coupled to a 
fluctuating gauge field $(a_0, {\bm a})$, where the temporal 
$a_0(r,\tau)$ enforce the local occupancy constraints while the spatial 
components represent the relevant fluctuations of 
$t_{rr'}(\tau) \approx |t| e^{i a_{rr'}(\tau)}$.
The corresponding continuum theory 
(``relativistic electrodynamics in a metal'') was studied in
Refs.~\onlinecite{Reizer, PALee, Polchinsky, Altshuler, Nayak, YBKim, Senthil}, and we will quote some results in the next section.

\vskip 2mm
In Appendix~\ref{app:props}, we study long-distance properties of the 
Gutzwiller-projected wave function in some detail.
As mentioned in the appendix, this wave function may be not sufficient 
to capture the long wavelength behavior of the actual phase, 
since the projection treats only the $a_0$ fluctuations, but does not 
include the fluctuations of $a_{rr'}$, while the latter are crucial in 
the effective theory.\cite{Reizer, PALee, Polchinsky, Altshuler, Nayak, YBKim, Senthil}
This is pointing a possible limitation of the projected wave function 
approach for the spinon-gauge system.
We still expect that the variational study of the previous section
gets the crude energetics correctly in the ring exchange model.
This is also what we expect from the mean field treatment, and
leads us to propose the effective spinon-gauge theory.
A finer numerical application likely requires more advanced techniques,
perhaps in the spirit of Ref.~\onlinecite{Imada} for the triangular
Hubbard model.  It would be interesting for example to look for the 
$2k_F$ signature\cite{Altshuler} in the more elaborate work of 
Ref.~\onlinecite{Imada}, which may be a more accurate realization
of the spinon-gauge ground state.

\section{Application to possible spin liquid state in 
$\kappa$-${\rm (ET)_2 Cu_2 (CN)_3}$}
\label{sec:exp}
We now discuss possible spin liquid state in the organic compound \ET,
which is insulating and shows no magnetic order down to the lowest 
experimental temperatures.  It is believed\cite{Imada, Kanoda, McKenzie} 
that the conducting layer of this material is well described by a 
single-band triangular lattice Hubbard model at half-filling with 
$t/U \simeq 1/8$ and only small hopping anisotropy of about $6\%$. 

Unlike the square lattice case, for the half-filled triangular lattice 
we expect a metallic phase for large enough $t/U$.
Reference~\onlinecite{Imada} estimates the metal-insulator
transition to occur at $(t/U)_{\rm MI} \simeq 1/5$, so the \ET 
material is on the insulating side.
Using an elaborate numerical technique, Ref.~\onlinecite{Imada} finds a 
nonmagnetic insulator in this regime.  
We want to develop some picture of this state.

The ideology we pursue here is that the insulating phase can
be described by an effective spin model.  Since the system
is close to the metal-insulator transition, it is not
enough to stop at two-spin exchange interactions.
Starting with the Hubbard model, the effective Hamiltonian 
to order $t^4/U^3$ was obtained in Ref.~\onlinecite{McDonald}.
Specialized to the triangular lattice, the spin Hamiltonian
reads
\begin{eqnarray}
\label{Heff}
\hat{H}_{\rm eff} = \hat{H}_{\rm ring}[J_2, J_4]
+ \sum_{\la\!\la ij \ra\!\ra} J^{\prime\prime} {\bm S}_i \cdot {\bm S}_j
+ \sum_{\la\!\la\!\la ij \ra\!\ra\!\ra} J^{\prime\prime\prime} 
{\bm S}_i \cdot {\bm S}_j ~.
\end{eqnarray}
Here $H_{\rm ring}$ is the ring exchange Hamiltonian Eq.~(\ref{Hring})
with $J_2 = (1-32t^2/U^2) 2t^2/U$, $J_4 = 20t^4/U^3$.
The effective Hamiltonian has additional Heisenberg exchanges
$J^{\prime\prime} = -16t^4/U^3$ between second neighbors
(separated by a distance $\sqrt{3}$) and 
$J^{\prime\prime\prime} = 4t^4/U^3$ between third neighbors
(separation $2$ lattice spacings).  Our grouping of the terms
in the effective Hamiltonian is intended to make it look as close as 
possible to the ring exchange model studied in the previous sections.

For the \ET compound, we estimate $J_4/J_2 \simeq 0.3$,
which puts the ring exchange model into the proposed spinon Fermi sea
regime.  Further neighbor interactions not included in the
$J_2$-$J_4$ model do not modify this result,
even though $J^{\prime\prime}$ and $J^{\prime\prime\prime}$
are roughly of the same magnitude as $J_4$.  This stability is because 
the corresponding further neighbor spin correlations are small
in the spin liquid regime.

\begin{figure}
\centerline{\includegraphics[width=\columnwidth]{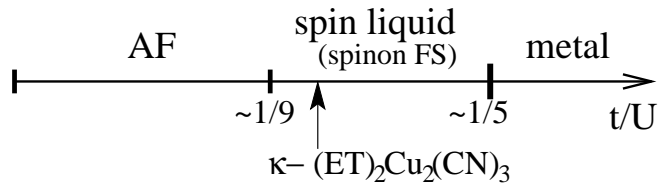}}
\vskip -2mm
\caption{Proposed phase diagram for the triangular lattice Hubbard
model.  The present study is based on the effective spin Hamiltonian
Eq.~(\ref{Heff}) and applies only to the insulating regime expected for 
$t/U \lesssim 1/5$ from Ref.~\onlinecite{Imada}.
Close to the metal-insulator transition, we propose the spin liquid 
state with spinon Fermi surface.  For smaller $t/U \lesssim 1/9$, 
the best state is AF ordered.  
The \ET compound has $t/U \simeq 1/8$
}
\label{material_phased}
\end{figure}

To proceed more systematically, we repeat the variational study with 
the effective Hamiltonian~(\ref{Heff}).
The resulting phase diagram is shown in Fig.~\ref{material_phased}
in terms of the Hubbard model parameter $t/U$.
From this study, we propose that the insulating ground state is the 
antiferromagnet for $t/U \lesssim 1/9$ (this corresponds roughly to 
the ring exchange parameter $J_4/J_2 \approx 0.2-0.25$).
For larger $t/U$, our best trial state is essentially the projected
Fermi sea state, and the variational $\Delta$ (which can be used to 
improve the trial energy slightly) is small already at the transition 
from the AF state.
In the same figure, we also indicate the metallic phase expected 
for $t/U \gtrsim 1/5$.

It should be emphasized that we do not treat either Hamiltonian 
Eq.~(\ref{Heff})~or~(\ref{Hring}) as more realistic or less realistic, 
particularly since we are dealing with the system near the 
metal-insulator transition.  
The above variational study with ${\hat H}_{\rm eff}$ is presented 
primarily to illustrate that our results are not destabilized by 
making the Hamiltonian `more realistic'.
We expect that our main prediction for the spin liquid state
close to the metal-insulator transition is robust, since the
proposed Gutzwiller-projected Fermi sea state is even more favored by 
including further effects of the electron kinetic energy.
Also, the results of Ref.~\onlinecite{Imada} give us some indication
on the stability of the proposed state, since that study is building
up on free-fermion states.

\subsection{Physical properties in the spin liquid phase with 
spinon Fermi surface}
The effective description of the proposed phase has spinon Fermi sea 
coupled to a dynamically generated gauge field.  It has been argued
\cite{Reizer, PALee, Polchinsky, Altshuler, Nayak, YBKim, Senthil}
that this spinon-gauge system is described by a nontrivial fixed point 
and shows unusual behavior, which can be tested in experiments.
Below, we list some thermodynamic properties of this Mott 
insulator.  This phase is in some sense the closest one can get to the 
Fermi liquid while remaining a charge insulator, and shares some 
properties with the metal due to the presence of the spinon Fermi 
surface, but also has some `non-Fermi liquid' properties.

\vskip 1mm
Thus, {\bf spin susceptibility} is expected to approach a constant
as temperature $T$ goes to zero:
\begin{equation}
\chi_{\rm spin}(T \to 0) \sim \mu_B^2 \nu_0 ~.
\end{equation}
This is a consequence of having gapless spinon excitations over the 
entire Fermi surface and is in fact observed in \ET.\cite{Ashvin_thanks}
Here, $\nu_0$ is the density of states at the ``Fermi surface'' in the 
spinon band structure determined by the spinon ``hopping amplitude'' 
$t_{\rm spinon}$.  The latter is set by the Heisenberg exchange energy 
$t_{\rm spinon} \sim J$ and is different from the bare electron 
hopping amplitude $t_{\rm el}$ (remember that $J \sim t_{\rm el}^2/U$).
For the triangular lattice at half-filling, we have 
$\nu_0 = 0.28 / t_{\rm spinon}$ per triangular lattice site and 
including spin.
Ref.~\onlinecite{Kanoda} reports $\chi=2.9 \times 10^{-4}$ emu/mol
at low temperatures, from which we estimate 
$t_{\rm spinon} \approx 350$K.
This compares favorably with the reported magnitude of the 
Heisenberg exchange $J=250$K.

\vskip 1mm
{\bf Specific heat}, on the other hand, is expected to show non-Fermi 
liquid behavior
\begin{equation}
C  \sim k_B \nu_0 \, t_{\rm spinon}^{1/3} (k_B T)^{2/3} ~.
\end{equation}
This is written to contrast with the Fermi liquid $\sim T$ behavior,
and means that the spin entropy in this charge insulator is in fact 
larger than in the metallic state at low temperature.  
This is very different from the antiferromagnet or gapped spin liquid 
insulators which have low spin entropy.  In particular, the finite 
temperature first-order transition line between the proposed spin liquid 
and the metallic state is expected to bend towards the metallic state 
with increasing temperature:\cite{Senthil_thanks}
\begin{equation}
p_{\rm MI}(T) - p_{\rm MI}(0) \sim T^{5/3} ~.
\end{equation}
In the last formula, $p$ is an applied pressure which drives the 
insulator to metal transition.\cite{Kanoda, Komatsu}
This tendency is actually observed in the \ET 
material.\cite{Kanoda_kitp}

\section{Conclusions}
In summary, we considered the spin-1/2 ring exchange model on
the triangular lattice from the variational perspective
and identified the instability of the antiferromagnetically ordered
state towards spin liquid state in the regime of moderate ring exchange 
couplings.
Our best trial states become the Gutzwiller-projected Fermi sea state
for larger $J_4$.
Despite the limitations of the variational approach, it is hoped that 
the present work may give complimentary information and useful guidance 
for understanding the exact diagonalization results.

We also studied the effective spin Hamiltonian appropriate for 
describing charge insulator states of the triangular lattice 
Hubbard model.  
The effective Hamiltonian includes Heisenberg exchanges as well as 
ring exchanges, and so is close to the considered ring exchange model.  
The study is motivated by the tentative spin liquid state in the 
\ET compound, which is modeled by the triangular lattice Hubbard model 
in the vicinity of the metal-insulator transition.
We find that upon including the ring exchanges but well in the 
insulating regime, the antiferromagnet gives way to the spin liquid 
state which is essentially the projected Fermi sea state.
In view of this finding, we propose that the effective description of 
the nonmagnetic insulator phase has Fermi sea of spinons
coupled to the dynamically generated gauge field.
This spin liquid phase features a number of unusual properties 
which can be looked for in experiments.  
It would be very exciting if this remarkable state is indeed 
realized in the \ET material.

\acknowledgements
The author has benefited from many useful discussions with
M.~P.~A.~Fisher, V.~Galitski, P.~Nikolic, and A.~Vishwanath,
and is especially grateful to T.~Senthil for motivating 
this problem and sharing many insights throughout the course of 
the study.  This work was started at MIT and supported by NSF grants 
DMR-0213282 and DMR-0201069.  The work at KITP is supported through 
NSF grant PHY-9907949.

\appendix

\section{Properties of the projected Fermi sea wave function}
\label{app:props}
\begin{figure}
\centerline{\includegraphics[width=\columnwidth]{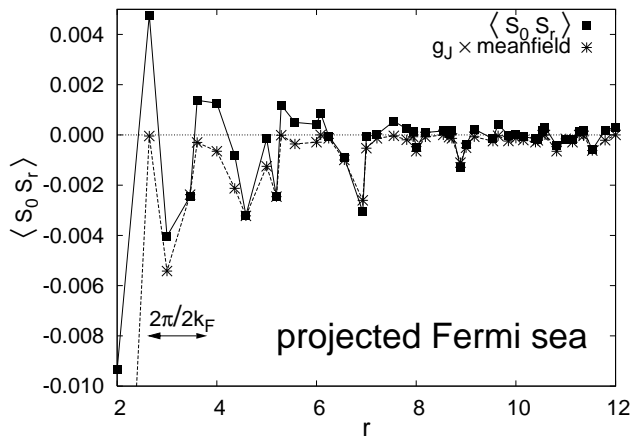}}
\vskip -2mm
\caption{Spin correlation in the projected Fermi sea state.
Measurements are done on a $24 \times 24$ triangular lattice.
The mean field wave function is constructed for periodic boundary 
conditions and excluding the zero momentum single particle state 
in order to avoid Fermi surface points while satisfying 
the lattice rotation symmetry for the finite system 
(this does not affect the long distance properties of the 
wave function which is our focus here).
Note the oscillating character of the correlation 
(with the period $\approx 2\pi/(2k_F)$, $k_F \approx 2.69$).
Also note that the renormalized mean field roughly reproduces the 
overall magnitude of the correlations.
}
\label{spin_corr_U1A}
\end{figure}

We describe some properties of the projected Fermi sea state.
Figure~\ref{spin_corr_U1A} shows spin correlations in the
projected wave function and also in the free fermion 
state before the projection.
In the free fermion state, the spin correlation behaves as
$-\cos^2(k_F r - 3\pi/4)/r^3$ at large distances, which oscillates 
with the wave vector $2k_F$ while always staying negative.
To facilitate the comparison, Fig.~\ref{spin_corr_U1A} shows 
the mean field correlations in the specific finite system
(the finite size effects are fairly large because of the
gaplessness over the Fermi surface).
We observe that the effect of the projection is not strong:
For the range studied, the mean field result multiplied by the 
Gutzwiller enhancement factor $g_J=4$ gives a reasonable approximation 
for the actual correlation.\cite{Zhang}
After the projection, the correlation function now swings to positive
values as well, but the overall magnitude is roughly
captured by the simple renormalization factor.

We also studied spin chirality correlations (not shown), and found
that these are very small beyond few lattice spacings.

The effective theory of the proposed phase has Fermi sea of spinons 
coupled to a dynamically generated gauge field.
\cite{Reizer, PALee, Polchinsky, Altshuler, Nayak, YBKim}
The measured spin correlations in the projected wave function
represent some puzzle in this respect:  
Ref.~\onlinecite{Altshuler} predicts that the spin structure factor is 
singularly enhanced near $2k_F$ in the spinon-gauge system.  
We find that in the projected wave function the structure factor 
remains finite throughout the Brillouin zone and that the overall 
rate of decay of spin correlations is roughly the same as in the 
free fermion state.  
One possible source of this difference is that the projected
wave function has fixed $t_{rr'}$ and therefore does not include the
fluctuations of the spatial components of the gauge field;
only the temporal component is ``included'' by the projection.
This is a limitation of the projected wave function 
approach for the spinon-gauge system.



\begin{thebibliography}{10}
\bibitem{Senthil_thanks}
These results were pointed out to the author by T. Senthil.
\end{thebibliography}
\end{document}